\documentclass[aps,reprint]{revtex4-1}

\usepackage{amsmath}
\usepackage{amsfonts}
\usepackage{bbm}
\usepackage{graphicx}
\graphicspath{{./}{../graphics/}}
\usepackage[dvipdfmx]{hyperref}

\begin{document}
\hfill{ NCTS-TH/1710, WU-HEP-17-11}

\title{On the Flux Vacua in F-theory Compactifications} 

\author{Yoshinori Honma${}^{1}$}
\email[]{yoshinori.honma255@cts.nthu.edu.tw}
\author{Hajime Otsuka${}^{2}$}
\email[]{h.otsuka@aoni.waseda.jp}
\affiliation{${}^1$ National Center for Theoretical Sciences,
National Tsing-Hua University, Hsinchu 30013, Taiwan \\
${}^2$ Department of Physics, Waseda University, Tokyo 169-8555, Japan}


\begin{abstract}
We study moduli stabilization of the F-theory compactified on
an elliptically fibered Calabi-Yau fourfold. Our setup is based on
the mirror symmetry framework including brane deformations.
The complex structure moduli dependence of the resulting 4D
${\cal{N}}=1$ effective theory is determined by the associated
fourfold period integrals. By turning on appropriate $G$-fluxes,
we explicitly demonstrate that all the complex structure moduli
fields can be stabilized around the large complex structure point
of the F-theory fourfold.
\end{abstract}

\maketitle

\section{Introduction}
\label{sec:intro}

String theory compactifications to four dimensional spacetime
provide a multitude of massless scalar fields. Unless these
extra moduli fields are stabilized, one cannot predict anything
for low energy physics including gravity. Moreover, the recent
observational data for the acceleration of the universe
motivated us to construct de Sitter vacua from a UV-complete
quantum theory of gravity. Under these circumstances, the
moduli stabilization and comprehensive study of flux vacua
have become one of the major topics in string theory. 

In the moduli stabilization, determination of the scalar potential of 
4D ${\cal{N}}=1$ effective theories arising from spacetime compactifications
is of particular interest. In the language of 4D ${\cal{N}}=1$
supersymmetry, there are two kinds of contributions to the scalar
potential of moduli fields, namely the K$\ddot{{\textrm{a}}}$hler
potential and the superpotential. The main problem of string compactifications
is how to derive these quantities quantum mechanically from the geometry of
internal compact spaces.

On the other hand, the mirror symmetry in string theory is known to be a useful
tool to understand exact properties of moduli fields of geometries as first
demonstrated to the quintic Calabi-Yau threefold in \cite{Candelas:1990rm}. As has been
explicitly performed in the literature, mirror symmetry can be applied to consider
the closed string moduli stabilization. Inclusion of open string sector in the presence of
the brane for the compact Calabi-Yau manifolds was initiated in \cite{Walcher:2006rs} and
has been subsequently applied in many contexts. In this framework, a brane is fixed on a
specific submanifold and the system does not have a continuous open string moduli
dependence. This means that the effective superpotential due to the wrapped branes
cannot be evaluated from this kind of undeformed setup. 

For the case of compact Calabi-Yau threefolds, the inclusion of brane deformations was
first carried out in \cite{Jockers:2008pe}. By using a Hodge theoretic approach, they
computed the brane superpotential depending on both open and closed string moduli. Thereafter, alternative and
more efficient methods to evaluate the brane superpotential has been constructed (see \cite{Alim:2009bx,
Grimm:2009ef,Jockers:2009ti,Grimm:2010gk} for details). Remarkably, these generalizations 
have led to a duality between open string on a threefold with branes and closed string on a
fourfold without branes, which can be naturally incorporated into the framework of the F-theory \cite{Vafa:1996xn}.
 
The F-theory conjecture implies that the physics of Type IIB string
compacitifications with branes on a complex three-dimensional
K$\ddot{{\textrm{a}}}$hler manifold can be encoded in the geometry of
an elliptically fibered Calabi-Yau fourfold. In contrast to various string
compactifications on Calabi-Yau threefolds, the moduli stabilization of F-theory
has not been fully established.
The aim of this work is to fill this gap by utilizing the mirror symmetry techniques to study the
F-theory vacua in the large complex structure limit, where the dynamics of moduli fields has not been
investigated explicitly. For other earlier attempts in a similar spirit, see \cite{Becker:1996gj,Dasgupta:1999ss,Gukov:1999ya} initiated the M-theory and F-theory compactifications with $G_4$ fluxes, \cite{Denef:2005mm} investigated the orientifold limit \cite{Sen:1996vd,Sen:1997gv} of F-theory and
\cite{Berglund:2005dm} based on the K3$\times$ K3 backgrounds.

\section{F-THEORY COMPACTIFICATIONS INCLUDING FLUXES}
\label{sec:F-COM}

First we describe basic ingredients for spacetime
compactification in the F-theory framework. For more details,
we refer the reader to \cite{Denef:2008wq}.

\subsection{F-theory on Calabi-Yau Fourfolds}
\label{sec:FonCY4}

Let us consider a class of 4D ${\cal{N}}=1$ effective theories arising from
F-theory compactified on the elliptically fibered Calabi-Yau fourfold
$X_4 \rightarrow B_3$. Here, $B_3$ is a complex three-dimensional
K$\ddot{{\textrm{a}}}$hler base space with positive curvature. This setup
can be also regarded as a Type IIB string theory compactified on $B_3$
with an axio-dilaton which varies over $B_3$ holomorphically.

In the F-theory perspective, the K$\ddot{{\textrm{a}}}$hler potential for 
complex structure moduli fields in 4D ${\cal{N}}=1$ effective theories can
be represented by
\begin{align}
K=-\ln{\int_{X_4}} \Omega \wedge \overline{\Omega},
\label{Kahp}
\end{align}
where $\Omega$ denotes a holomorphic $(4,0)$-form on $X_4$.
Here and in what follows, we have adopted the reduced Planck unit
$M_{\rm Pl}=2.4\times 10^{18}\,{\rm GeV}=1$. 
It is also well-known that F-theory admits a superpotential of the form
\begin{align}
W=\int_{X_4} G_4 \wedge \Omega,
\end{align} 
in the presence of non-zero four-form fluxes $G_4$. This expression is
inherited from a duality between F-theory and M-theory \cite{Becker:1996gj,Sethi:1996es,Haack:2001jz,Denef:2008wq}.
To guarantee the compactness of a background, the above $G_4$ fluxes
are required to satisfy the tadpole cancellation condition given by 
\begin{align}
\frac{\chi}{24}=n_{\rm{D3}}+\frac{1}{2}\int_{X_4} G_4 \wedge G_4,
\end{align}
where $\chi$ is the Euler characteristic of $X_4$ and $n_{\rm{D3}}$
denotes the total charge of the space-time filling D3-branes. 

Note that generically the variations of $\Omega$
in Calabi-Yau fourfolds do not completely span $H^4(X_4)$
and only its primary horizontal subspace given by
\begin{align}
H^4_H (X_4, {\mathbb{C}}) 
= H^{4,0} \oplus H^{3,1} \oplus H^{2,2}_H \oplus H^{1,3} \oplus H^{0,4}
\end{align}
can contribute to the mirror symmetry calculations \cite{Greene:1993vm}.
Here $H^{2,2}_H$ denotes the elements in $H^{2,2}$ which arise from the
second derivatives of $\Omega$ with respect to the complex structure
moduli of $X_4$. Correspondingly, the middle dimensional homology basis
is also restricted to lie in the primary horizontal subspace of $H_4 (X_4)$.

Concerning the dynamics of the K$\ddot{{\textrm{a}}}$hler moduli fields in 
effective theories, it is quite challenging to elicit exact interactions from
internal geometry. One of the difficulties in this determination is due to the
lack of understanding about the quantum moduli space of hypermultiplets
(see \cite{Alexandrov:2013yva} for recent developments) 
and the possibility of its consistent reduction to the 4D ${\cal N}=1$ supergravity 
formulation in general region of the moduli space. 
Here we simplify the situation and only add a classical term $-2 \ln{\cal{V}}$ to
the K$\ddot{{\textrm{a}}}$hler potential and assume that the radius of the 
background manifold is sufficiently large so that the classical K$\ddot{{\textrm{a}}}$hler moduli space 
has a no-scale structure~\cite{Giddings:2001yu}. 
This additional term corresponds to the volume of a background and is
in general a nontrivial function of the K$\ddot{{\textrm{a}}}$hler moduli and the
mobile D3-brane moduli \cite{Denef:2008wq}. We assume that $\cal{V}$ will be
stabilized at a particular constant after the complex structure moduli stabilization,
as first demonstrated in \cite{Kachru:2003aw}.

\subsection{Moduli dependence}
\label{sec:mirror}

First we will describe general aspects of complex structure moduli dependence
in F-theory compactifications. More concrete expressions based on a fixed
background will be presented in the next subsection.

For a Calabi-Yau fourfold $X_4$ with $h^{3,1}(X_4)$ complex structure
moduli, the period integrals of holomorphic $(4,0)$-form $\Omega$ defined by
\begin{align}
\Pi_i = \int_{\gamma^i} \Omega
\label{4peri}
\end{align}
encode a closed string moduli dependence of the system. Here, $\gamma^i$ with
$i=1, \ldots , h^4_H (X_4)$ denote a basis of primary horizontal subspace
of $H_4 (X_4)$. In terms of these fourfold periods, the K$\ddot{{\textrm{a}}}$hler
potential for the complex structure moduli (\ref{Kahp}) can be written as
\begin{align}
K=-\ln{( \sum_{i,j} \Pi_i \eta^{ij} \overline{\Pi}_j)},
\label{KbyP}
\end{align}
where we have introduced a moduli independent intersection matrix $\eta^{ij}$
and a dual basis $\hat{\gamma}^i$ in $H^4_H (X_4)$ as
\begin{align}
\eta^{ij} = \int_{X_4} \hat{\gamma}^i \wedge \hat{\gamma}^j,
 \ \ \ \ \int_{\gamma^i} \hat{\gamma}^j = \delta^{ij}. 
\end{align}

Now we consider turning on a class of $G_4$ fluxes whose integer quantum
numbers are given by
\begin{align}
n_i = \int_{\gamma^i} G_4.
\label{fluq}
\end{align}
These fluxes generate a superpotential for the complex structure moduli
of the form
\begin{align}
W= \sum_{i,j} n_i \Pi_j \eta^{ij}.
\end{align}

Note that our choice of $G_4$ fluxes (\ref{fluq}) only involved with
$H^4_H (X_4)$. In general, there exists additional contributions to the
system from other subspaces of $H^4(X_4)$ (see e.g. \cite{Braun:2014xka}). 
More rigorous treatment for the couplings arising from these remaining
$G_4$ fluxes would be indispensable for studying the stabilization of 
K\"ahler moduli fields.

\subsection{Topological data}
\label{sec:data}

As a simplest example of a fourfold $X_4$, we consider an elliptically
fibered Calabi-Yau fourfold $X_c^*$ which has been constructed in \cite{Alim:2009bx}
from the quintic Calabi-Yau threefold with one toric brane (see also \cite{Jockers:2009ti}).
For details about F-theory fourfold construction, we refer the reader to \cite{Grimm:2009ef} where a general
analysis about the mirror pairs for the elliptic Calabi-Yau fourfolds has
been clarified. Note that not every Calabi-Yau threefold can be uplifted to the
consistent F-theory fourfold background in these prescriptions. As mentioned in \cite{Grimm:2009ef},
the existence of an elliptic fibration structure in the mirror of the underlying threefold is
crucial for the fourfold uplifting.

The period integrals (\ref{4peri}) for the fourfold $X_c^*$ have been obtained in \cite{Alim:2009bx,Jockers:2009ti}
by using toric geometry techniques and the result is
\begin{align}
\begin{split}
\Pi_1&=1, \ \Pi_2=z, \ \Pi_3=-z_1, \ \Pi_4=S, \\
\Pi_5&=5Sz, \ \Pi_6=\frac{5}{2}z^2, \ \Pi_7=2z_1^2, \ \Pi_8=-\frac{5}{2}Sz^2-\frac{5}{3}z^3, \\
\Pi_9&=-\frac{2}{3}z_1^3, \ \Pi_{10}=-\frac{5}{6}z^3, \ \Pi_{11}=\frac{5}{6}Sz^3+\frac{5}{12}z^4-\frac{1}{6}z_1^4,
\end{split}
\label{peri}
\end{align}
where we have ignored further possible corrections to the leading interactions. The complex structure moduli
of the fourfold $z,z-z_1,S$ are originated from a bulk quintic modulus, a brane modulus and the axio-dilaton
in Type IIB description, respectively.

Note that our definition for the complex structure moduli fields $\{ z \}$ deviates from the standard convention
also used in \cite{Alim:2009bx}, where the classical periods are expressed by logarithmic functions of
the complex structure deformations. We redefined a logarithm of a standard complex
structure modulus as a new single modulus just for later convenience.

The topological intersection matrix has been also clarified in \cite{Alim:2009bx} as
\begin{align}
\eta=\begin{pmatrix}
0& 0 & 0 & 0 & 1 \\
0 & 0 & 0 & I_3 & 0 \\
0 & 0 & \widetilde{\eta} & 0 & 0 \\
0 & I_3 & 0 & 0 & 0 \\
1 & 0 & 0 & 0 & 0
\end{pmatrix}, \ \ 
\widetilde{\eta}=\begin{pmatrix}
0& \frac{1}{5} & 0 \\[5pt]
\frac{1}{5} & \frac{2}{5} & 0 \\[5pt]
0 & 0 & -\frac{1}{4}
\end{pmatrix},
\label{intm}
\end{align}
and the Euler characteristic of the background is given by $\chi (X_c^*)=1860$.

\section{ILLUSTRATIVE EXAMPLE OF MODULI STABILIZATION}

\subsection{Effective theory for moduli fields}
\label{sec:eft}

Here we describe the explicit form of the 4D ${\cal{N}}=1$ effective potentials
for moduli fields arising from F-theory compactified on $X_c^*$.
Substituting the fourfold data (\ref{peri}) and (\ref{intm}) into (\ref{KbyP}),
one can easily check that the K$\ddot{{\textrm{a}}}$hler potential for moduli
fields takes a form
\begin{align}
K=-\ln{\left[ -i(S-\overline{S}) \right]} -\ln{\widetilde{Y}}-2\ln{\cal{V}},
\end{align}
where
\begin{align}
\widetilde{Y}=\frac{5i}{6}(z-\bar{z})^3+\frac{i}{S-\overline{S}}
\left( \frac{5}{12}(z-\bar{z})^4 -\frac{1}{6}(z_1-\bar{z}_1)^4 \right),
\end{align}
and we have added the simplified K$\ddot{{\textrm{a}}}$hler moduli sector.
Note that our simplification for K$\ddot{{\textrm{a}}}$hler moduli fields
does not affect the later discussion about the vacuum structure of F-theory 
compactifications, as long as the masses of K\"ahler moduli fields are significantly
smaller than the other moduli fields. 

Similarly, the superpotential can be written as
\begin{align}
W &= n_{11}+n_{10}S+n_8 z+n_6 S z+\frac{5}{2}\left( \frac{n_5}{5}+\frac{2 n_6}{5} \right) z^2 \nonumber \\
& -\frac{5n_4}{6} z^3-n_2 \left( \frac{5}{2}Sz^2+\frac{5}{3}z^3 \right)-n_9 z_1-\frac{n_7}{2}z_1^2 \nonumber \\
& -\frac{2n_3}{3}z_1^3+n_1 \left( \frac{5}{6}Sz^3+\frac{5}{12}z^4-\frac{1}{6}z_1^4 \right),
\label{W}
\end{align}
and the tadpole cancellation condition takes a form
\begin{align}
\frac{1860}{24} =& \ n_{\rm{D3}}+n_1 n_{11}+n_2 n_8+n_3 n_9+n_4 n_{10} \nonumber \\
& +\left( \frac{n_5+n_6}{5}\right) n_6-\frac{n_7^2}{8}.
\label{tadcan}
\end{align}
Obviously, $n_7$ must be $2+4k$ with $k \in {\mathbb{Z}}$ in order to satisfy 
the condition (\ref{tadcan}) while preserving the integrality of flux quanta.
In a similar reason, $n_5+n_6$ or $n_6$ is constrained to be $5k'$ with $k' \in {\mathbb{Z}}$.

\subsection{F-theory flux vacua}
\label{sec:exam}

Let us study the extremal conditions of moduli fields $\Phi^I = (z,z_1,S)$. 
The $F$-term scalar potential of our 4D ${\cal{N}}=1$ effective theory
for moduli fields has a form
\begin{align}
V = e^K \left( K^{I \bar{J}} D_I W D_{ \bar{J}} \overline{W} \right),
\end{align}
where $D_{I} = \partial_I + (\partial_I K)$ and
$K^{I \bar{J}}$ is the inverse of the K$\ddot{{\textrm{a}}}$hler metric
given by $K_{I \bar{J}} = \partial_I \partial_{\bar{J}} K$.
Note that the no-scale structure of the K$\ddot{{\textrm{a}}}$hler moduli
fields \cite{Giddings:2001yu} is preserved at the classical level 
and a term proportional to $-3|W|^2$ in the standard 4D ${\cal N}=1$ formula is canceled.
Here we also define
\begin{align}
F^I \equiv K^{I \bar{J}} D_{\bar{J}} \overline{W},
\end{align}
for later convenience.
In this notation, the extremal conditions for moduli fields become
\begin{align}
\begin{split}
&e^{-K} \frac{\partial V}{\partial {\overline{\Phi}^{\bar{I}}}} = 
\left[ K_{\bar{I}}K_{J \bar{L}}-\partial_{\bar{I}}
K_{J \bar{L}}+K_J K_{\bar{I} \bar{L}} \right] F^J \overline{F}^{\bar{L}} \\
& \ +\overline{F}^{\bar{J}} \overline{W}_{\bar{J}\bar{I}}+(K_{\bar{J} \bar{I}}
-K_{\bar{J}}K_{\bar{I}}) \overline{F}^{\bar{J}}\overline{W}+K_{J \bar{I}}F^J W =0.
\end{split}
\end{align}

Here we focus on the self-dual $G_4$ fluxes satisfying
\begin{align}
G_4=*_{X_4} G_4,
\end{align}
which correspond to the imaginary self-dual 
three-form fluxes in Type IIB compactifications. 
In our model, imposing the self-duality condition is equivalent to set
$n_2=n_3=n_4=n_8=n_9=n_{10}=0$.
In this setup, our ${\cal{N}}=1$ effective theory
has a solution to the $F$-term conditions $F^I=0$, where the scalar potential
becomes zero and the values of the moduli fields
are fixed as
\begin{align}
\begin{split}
{\textrm{Re}}z &={\textrm{Re}}z_1={\textrm{Re}}S=0, \\
{\textrm{Im}}z&=\left(\frac{6n_{11}}{5n_1}\right)^{1/4}\frac{2\sqrt{n_6}}
{(8n_6(n_5+n_6)-5n_7^2)^{1/4}},\\ 
{\textrm{Im}}z_1&=\left(\frac{30n_{11}}{n_1}\right)^{1/4}\frac{\sqrt{n_7}}
{(8n_6(n_5+n_6)-5n_7^2)^{1/4}}, \\
{\textrm{Im}}S &=\left(\frac{6n_{11}}{5n_1}\right)^{1/4}\frac{n_5}{\sqrt{n_6}
(8n_6(n_5+n_6)-5n_7^2)^{1/4}}.
\label{vacuum}
\end{split}
\end{align}

For example, there exists a Minkowski vacuum with $n_{\rm{D3}}=0$ in the following
choice of non-zero $G_4$ fluxes:
\begin{align}
n_1=1,n_5=15,n_6=10,n_7=2,n_{11}= 28.
\end{align}
The values of the moduli fields (\ref{vacuum}) in this vacuum are
\begin{align}
\begin{split}
{\textrm{Re}}z &={\textrm{Re}}z_1={\textrm{Re}}S=0, \\
{\textrm{Im}}z &\simeq 2.28, \ \ {\textrm{Im}}z_1 \simeq 1.14, \ \  {\textrm{Im}}S \simeq 1.71,
\end{split}
\end{align}
and the vacuum expectation value of the superpotential becomes $W \simeq -72.97$.
One can easily confirm that the mass eigenvalues of moduli fields are
positive definite as
\begin{align}
{\cal{V}}^{-2} (91.30, \ 35.05, \ 3.94, \ 2.96, \ 0.09, \ 0.07),
\end{align}
which means that all the complex structure moduli have been completely stabilized.

\vspace{-0.2pt}

\section{Conclusions}
\label{sec:conclusions}

It has been known that the effective superpotential and the axio-dilaton
dependence of Type IIB compactifications can be reformulated into a geometry
and fluxes in F-theory. Meanwhile, exact calculations in such a situation has been
studied in a framework of mirror symmetry with or without branes. In this work,
we have shown that topological data extracted by mirror symmetry techniques
can be directly applied to the F-theory compactifications. Especially we have
demonstrated that all the complex structure moduli can be stabilized around the large
complex structure point of F-theory fourfold.

Throughout this work, we have only focused on classical interactions of moduli fields.
This means that we have not fully utilized the power of mirror symmetry and further
quantum corrections to the effective couplings can be also easily calculated. It would be
interesting to study the vacuum structure of F-theory including these corrections, which
can be also computed as in \cite{Honma:2013hma}.

Moreover, it would be fascinating to check whether the K\"ahler moduli can be stabilized
as in the LARGE Volume Scenario \cite{Balasubramanian:2005zx} or the scenario
of the Kachru, Kallosh, Linde and Trivedi \cite{Kachru:2003aw}, once our treatment
of the K$\ddot{{\textrm{a}}}$hler moduli sector is extended.



\begin{acknowledgments}
  We would like to thank K.~Choi, C.~S.~Chu and H.~Hayashi
  for useful discussions and comments.
  Y.~H. was supported in part by the grant MOST-105-2119-M-007-018, 106-2119-M-007-019
  from the Ministry of Science and Technology of Taiwan.
  H.~O. was supported in part by Grant-in-Aid for Young Scientists (B) (No.~17K14303) 
  from Japan Society for the Promotion of Science. 
\end{acknowledgments}


\end{document}